# In-situ optical vector analysis based on integrated lithium niobate single-sideband modulators

Hanke Feng[1], Tong Ge[1], Yaowen Hu[2,3], Zhenzheng Wang[1], Yiwen Zhang[1], Zhaoxi Chen[1], Ke Zhang[1], Wenzhao Sun[1,4,5], and Cheng Wang[1*]

[1]*Department of Electrical Engineering & State Key Laboratory of Terahertz and Millimeter Waves, City University of Hong Kong, Kowloon, Hong Kong, China*
[2]*State Key Laboratory for Mesoscopic Physics and Frontiers Science Center for Nano-optoelectronics, School of Physics, Peking University, Beijing, China*
[3]*John A. Paulson School of Engineering and Applied Sciences, Harvard University, Cambridge, MA, USA*
[4]*City University of Hong Kong (Dongguan), Dongguan, China*
[5]*Center of Information and Communication Technology, City University of Hong Kong Shenzhen Research Institute, Shenzhen, China*

*cwang257@cityu.edu.hk

**Abstract:** Optical vector analysis (OVA) is an enabling technology for comprehensively characterizing both amplitude and phase responses of optical devices or systems. Conventional OVA technologies are mostly based on discrete optoelectronic components, leading to unsatisfactory system sizes, complexity, and stability. They also encounter challenges in revealing the on-chip characteristics of integrated photonic devices, which are often overwhelmed by the substantial coupling loss and extra spectral response at chip facets. In this work, we demonstrate a miniaturized OVA system for integrated photonics devices based on broadband single sideband (SSB) modulators on a thin-film lithium niobate (LN) platform. The OVA could provide a direct probe of both amplitude and phase responses of photonic devices with kHz-level resolution and tens of terahertz measurement bandwidth. We perform in-situ characterizations of single and coupled microring resonators fabricated on the same chip as the OVA, unfolding their intrinsic loss and coupling states unambiguously. Furthermore, we achieve the direct measurement of collective phase dynamics and density of states of the Bloch modes in a synthetic frequency crystal, by in-situ OVA of a dynamically modulated microring resonator. Our in-situ OVA system provides a compact, high-precision, and broadband solution for characterizing future integrated photonic devices and circuits, with potential applications ranging from optical communications, biosensing, neuromorphic computing, to quantum information processing.

## 1. Introduction

Accurately revealing the spectral information of optical devices and systems is of the utmost importance for comprehending performance characteristics, understanding the underlying physics, and discovering potential applications [1]. Optical vector analysis (OVA), in analogy to electrical vector analysis, is a powerful tool to perform such characterizations by capturing both amplitude and phase information [2-15]. Traditional OVA approaches primarily rely on optical interferometry [2] or modulation phase-shift techniques [3]. The necessity of laser wavelength scanning in these methods often leads to inadequate measurement resolutions on the MHz level, posing challenges in keeping pace with the fine spectral features in emerging optical devices (down to kHz level). Recently, OVAs based on the concept of microwave photonics (MWP) [16,17] have been proposed and demonstrated [4-14], by performing optical frequency scanning and signal measurement in the electrical domain through electro-optic (EO) modulation and photodetection. Benefiting from the ultrahigh-resolution frequency sweeping and accurate vector measurement ability in the electrical domain, MWP-based OVA technology is endowed with hyperfine resolution down to theoretically hertz level and the capability to extract both amplitude and phase information [7]. Nevertheless, all existing MWP-based OVA systems have been demonstrated either using all-discrete optical devices [4-12] or by combining silicon signal processing chips with traditional off-the-shelf modulators [13,14]. These approaches inevitably lead to increased system cost, bulkiness, and complexity, prohibiting the deployment in scenarios involving mobile and remote devices such as drones, autopilot vehicles, and satellites. In addition, all aforementioned demonstrations are ex-situ measurements where the test equipment and devices under test (DUT) are physically separated, which is prone to environmental fluctuations and de-embedding errors from the link components between the test instrument and DUTs. This is particularly problematic for measuring the spectral responses of on-chip photonic devices, which are often overwhelmed by substantial fiber-chip coupling losses. Moreover, the presence of undesired

Fabry-Perot interference caused by coupling facet reflection often introduces additional undesired amplitude and phase features, resulting in substantial uncertainties in the measurement results.

The recently emerged thin-film lithium niobate (LN) MWP platform [18] exhibits tremendous potential in resolving the aforementioned challenges and realizing integrated OVA systems with dramatically reduced size, weight, and power consumption (SWaP), owing to its unique EO properties and low optical loss characteristics [19,20]. To date, many key MWP building blocks have been demonstrated on the LN platform, including linear and broadband EO modulators [21-25], low-loss functional elements with various spectral responses [26-29], EO/Kerr comb sources [30,31], and on-chip stimulated Brillouin scattering (SBS) devices [32,33]. The excellent performances demonstrated in these building blocks could significantly enhance the MWP system metrics including bandwidth, noise figure, reconfigurability and spectral resolution [16]. Endeavors to further integrate these elements into chip-scale systems have led to integrated LN MWP systems with unparalleled performances, including ultrahigh-speed, low-power analog signal processors [18] and broadband multi-purpose photonic millimeter-wave radars [34]. In addition to the prospect of miniaturizing OVA systems, the excellent scalability of the LN platform also unlocks the possibility of seamlessly integrating the photonic DUTs with the OVA system on the same chip, enabling the much desired in-situ measurements for integrated photonics.

Here, we demonstrate such an in-situ MWP-based OVA system based on an LN photonic integrated circuit, capable of directly probing the amplitude and phase responses of integrated photonic devices. The integrated OVA system builds on an SSB modulator, which consists of a broadband phase modulator and a reconfigurable flat-top filter, for high-fidelity EO conversion. Fabricated from a 4-inch wafer-scale process, the device is capable of sweeping the modulated optical probe signal with a 20-dB sideband suppression ratio (SSR) and 50-kHz frequency resolution. The total measurement bandwidth could be extended up to tens of THz by shifting and stitching different measurement channels leveraging the excellent wavelength reconfigurability of our SSB modulator. Proof-of-concept in-situ measurements are carried out on single and coupled microring resonators fabricated on the same chip as the OVA, unveiling the intrinsic loss and coupling states of these systems that are often ambiguous with amplitude measurements only. Further performing in-situ OVA of an actively modulated microresonator enables, for the first time, directly probing the collective phase dynamics of the Bloch modes in a synthetic frequency crystal.

## 2. Results

Figure 1 shows the schematic illustration and working principle of our proposed in-situ OVA system. A continuous-wave optical carrier [$f_c$, (i)] is modulated by high-resolution scanning RF probe signals [$f_1 \sim f_2$, (ii)] using an on-chip EO phase modulator, which generates typical double sideband (DSB) signals with scanning optical frequencies in both sidebands (iii). The DSB signals are subsequently reshaped by an optical flat-top band-pass filter [blue curve in (iv)] that preserves only the carrier and one sideband, leading to SSB modulation (iv). The remaining SSB signals are divided into two branches, wherein one is directed towards the DUT, and the other goes through a bare optical waveguide with the same path length as a reference arm for calibrating the OVA system. After transmitting through the DUT, the scanning SSB probe signals capture the frequency-dependent amplitude and phase response of the DUT (using double microring resonators as an example, v). Afterwards, the SSB probe signals mix with the optical carrier at the photodetector, rendering high-resolution electrical signals that preserve both amplitude and phase information of the DUT thanks to the coherent nature of the entire process (vi) (see Methods). To facilitate in-situ measurements, various on-chip DUT, including passive [microring resonators, double-ring resonators, coupled-resonator optical waveguides (CROW)] and active devices (resonant EO comb generator), are fabricated and positioned next to the SSB modulators.

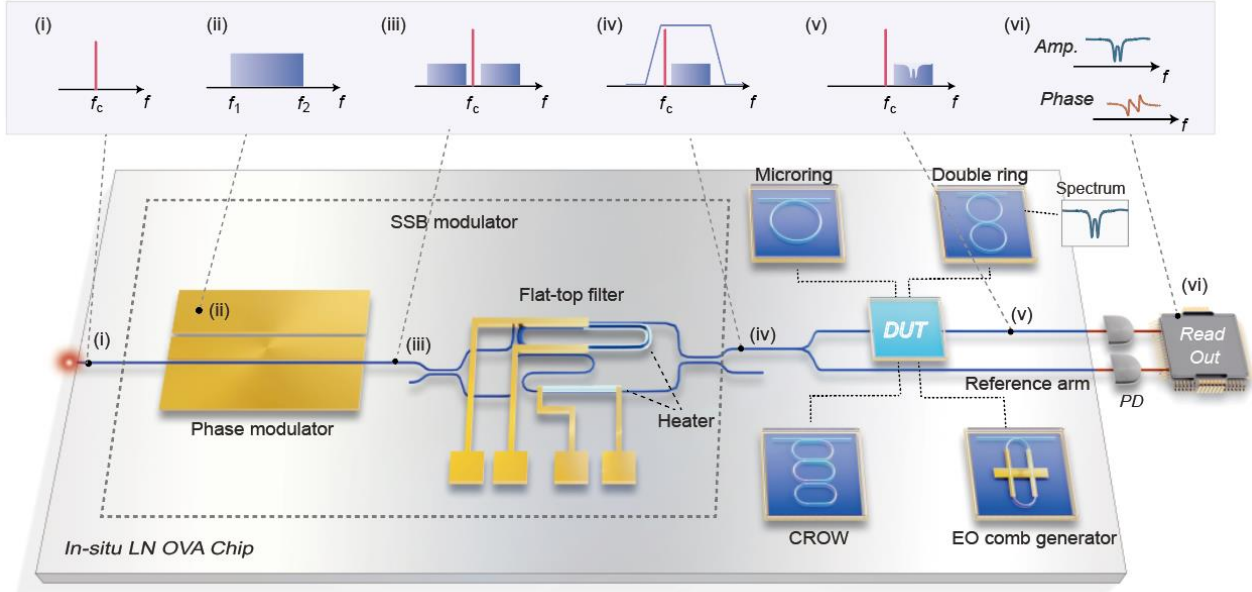

**Fig. 1. Schematic illustration and working principle of the integrated LN OVA system.** The SSB modulator consists of a phase modulator for signal up-conversion, and a tunable flat-top bandpass filter for single sideband suppression. Various passive and active DUTs are fabricated on the same chip as the SSB modulator for in-situ measurements. Insets (i-vi) schematically illustrate the spectra of optical and electrical signals at different locations of the chip. SSB: single sideband, DUT: device under test, CROW: coupled-resonator optical waveguides, Amp: amplitude.

The integrated LN OVA is fabricated on a 4-inch wafer-scale manufacturing platform using an ultraviolet (UV) stepper lithography system (see Methods). Figure 2a shows the false-color optical image of the LN SSB modulator, consisting of a high-performance phase modulator (EO bandwidth > 50 GHz) with advanced slotted-electrodes [25] and a tunable flat-top optical filter based on a ring-assisted Mach-Zehnder interferometer (RAMZI) configuration [35-37]. The left inset in Fig. 2a shows a photograph of a fabricated chip clamped by tweezers. Middle and right insets highlight details of the slotted electrodes and thermo-optic (TO) phase shifter through scanning electron microscope (SEM) images. The operation principle of the RAMZI flat-top filter is shown in Fig. 2b, where the top and middle panels plot the optical transmission and phase response respectively, of a microring resonator (light blue) and a Mach-Zehnder interferometer (MZI, dark blue), as functions of wavelength detuning [normalized by the free spectral range (FSR) of MZI]. To build a flat-top bandpass filter [36], the FSR of the microring resonator should equal half the FSR of the MZI, such that the extra phase induced by microring resonances always applies on the slope (i.e. quadrature point) of the sinusoidal transfer function of the MZI. By carefully engineering the coupling coefficient between the microring resonator and the bus waveguide to be $\kappa$=0.94 using a multi-mode interferometer (MMI) (see Methods), an ideal phase relationship between the microring and the MZI could be achieved, leading to a sharp transition from on to off states (or vice versa) near the quadrature points. As a result, the transmission spectrum is strongly modified from the typical sinusoidal curve of an asymmetric MZI to a "box-like" response with near-flat and periodically occurring pass and stop bands (bottom panel of Fig. 2b). In actual experiments, non-ideal coupling coefficients $\kappa$ could lead to undesirable phase relationship at the off-resonance points (dashed lines in top panel of Fig. 2c), resulting in ripples in the pass/stop bands (bottom panel of Fig. 2c) and compromised extinction ratios.

To take full advantage of the broad modulation bandwidths of LN modulators, we design FSRs of the microring resonator and the asymmetric MZI to be 120 GHz and 60 GHz, respectively, resulting in a flat pass-band of ~ 50 GHz. The resonant wavelength of the microring resonator is aligned to the quadrature point of the MZI by TO phase shifters fabricated in the vicinity of both the microring resonator and one arm of the MZI (Fig. 2a). The measured spectral response of the RAMZI filter in a representative channel is shown in Fig. 2d (grey), with a 210 dB/nm roll-off slope and a 20-dB extinction ratio. By locating the optical carrier at the edge of the pass band, SSB modulation could be achieved within a wide RF frequency range between 10~50 GHz, with measured optical SSR consistently at ~ 20 dB (Fig. 2d). Moreover, the pass band of our RAMZI filter could be flexibly shifted without affecting the flat-top nature, by simultaneously tuning the TO phase shifters on the microring and the MZI. Fig. 2e shows a series of measured flat-top bandpass profiles when the TO phase shifters are adjusted to set the relative phases of MZI at 0, $\pi/2$, $\pi$, and $3\pi/2$ (from top to bottom), with TO tuning efficiencies of 12 pm/mW for the microring resonator and 5 pm/mW for the MZI, respectively. The ability to shift the pass band over a full period unlocks the possibility of achieving an ultrawide measurement bandwidth by separately performing OVA within individual bands, and subsequently shifting and stitching the channels.

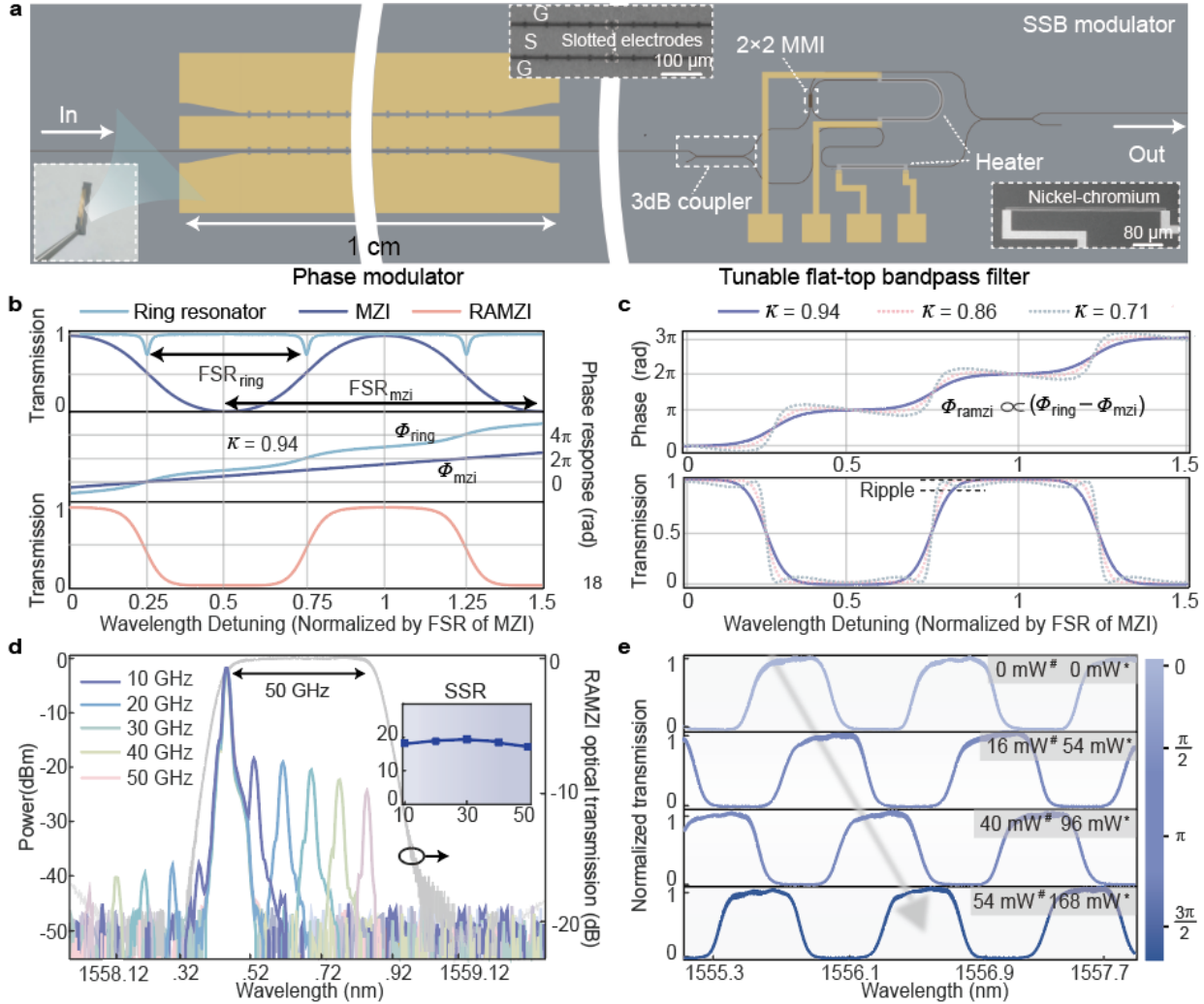

**Fig. 2 Principle and characteristics of LN SSB modulators. a** Optical image (false-color) of the LN SSB modulator, including a phase modulator and a RAMZI tunable flat-top filter. Left inset: photograph of the chip clamped by tweezers. Middle and right insets: scanning electron micrographs (SEM) of the slotted modulation electrodes and the TO phase shifter, respectively. **b** Principle of the RAMZI flat-top filter. Top and middle panels show the output optical amplitude transmission and phase response, respectively, of a microring resonator (light blue) and an unbalanced MZI (dark blue), as functions of normalized wavelength detuning (per FSR of MZI). By aligning the resonance notches to the quadrature points of the MZI and choosing a proper coupling state ($\kappa$=0.94), the additional phase shift from the microring resonator modifies the linear phase response of MZI, turning the full RAMZI transfer function to a "box-like" flat-top bandpass profile as shown in the bottom panel. **c** Simulated phase response and output transmission of the RAMZI filter at different coupling states of the microring resonator, showing undesired ripples in the pass/stop bands in the cases of non-ideal coupling coefficients. **d** Measured optical transmission spectrum of the RAMZI filter (grey) with a 50 GHz flat-top passband, a 210 dB/nm roll-off slope, and a 20-dB extinction ratio, enabling SSB modulation (color-coded output spectra) with 20-dB SSR within 10~50 GHz (inset). **e** Measured bandpass profiles when the relative phases of MZI are set at 0,

π/2, π, and 3π/2 (from top to bottom). # Heater power of ring resonator. * Heater power of MZI.

## 3. In-situ OVA of integrated passive devices

To implement the in-situ measurement of integrated photonic devices, we fabricate various DUTs next to the SSB modulators on the same LN chip and perform OVA using the experimental setup shown in Fig. 3a. High-resolution sweeping RF signals (10~50 GHz) from an electrical vector network analyzer (EVNA) are used to drive the SSB modulator and probe the optical properties of DUT following the working principle introduced above. The signals are converted back into the electrical domain using a high-speed photodetector and analyzed at the other port of the EVNA (see Methods). All demonstrations are performed using small-signal modulation to minimize measurement errors induced by higher-order sidebands [12]. The fabricated passive on-chip devices for in-situ OVA include a single-ring resonator, a double-ring coupled-resonator system, and a triple-ring CROW. These devices are fundamental building blocks for a variety of applications in integrated photonics, such as electromagnetically induced transparency (EIT) models [38], non-Hermitian photonics [39], and optical filters [40].

Figure 3b depicts the measured and fitted amplitude and phase responses of a single microring resonator, showing two optical modes (a fundamental and a high-order mode) simultaneously present in the measurement window with resonant frequencies separated by 6.2 GHz. The amplitude responses of the two resonances indicate notch depths of 4 dB and 11 dB, and full width at half maximum (FWHM) linewidths (in linear scale) of 500 MHz and 378 MHz, respectively. Importantly, the vector analysis adopted in our experiments allows us to directly probe the phase responses of these resonances and infer their coupling states (under/critical/over), which is typically not possible in an amplitude-only transmission spectrum measurement. In this particular case, we measure resonant phase jumps of 30 and 76 degrees for the two resonances, both of which correspond to under-coupling states. Further fitting the experimental results with theoretical model (dashed grey curves) provides unambiguous and quantitative analysis on the coupling coefficients of these two modes at $\kappa_1$=0.092 (high-order mode) and $\kappa_2$=0.093 (fundamental mode); as well as the intrinsic quality factors of $3.86\times10^5$ and $9.03\times10^5$, respectively. To validate the high resolution of our OVA system, the right panel of Fig. 3b shows the fine measurement results near one resonance notch, which is comprised of 40001 data points over a span of 2 GHz, resulting in a resolution of 50 kHz, currently limited by the linewidth of our tunable laser. The implementation of a narrow linewidth laser could in principle enable hyperfine resolution down to hertz-level [7,12].

For more complicated photonic systems like the double-ring resonator shown in Fig. 3c, our in-situ OVA system could also reliably capture the resonant phase jumps for both split modes, i.e. 40 and 78 degrees in this case. The fitting results reveal an initial resonance frequency difference($\Delta f$) of 0.457 GHz between the two ring resonators due to fabrication non-uniformity, a ring-bus coupling coefficient of $\kappa_1$=0.16, and a coupling coefficient between the two ring resonators of $\kappa_2$=0.03. We further conduct in-situ vector analysis of a triple-ring CROW system at the through port, as shown in Fig. 3d. Assuming the two ring-ring couplers ($\kappa_r$) and the two ring-bus couplers ($\kappa_b$) are both identical, our measurement and fitting results again lead to reliable estimation of the coupling coefficients, i.e. $\kappa_r$=0.26 and $\kappa_b$=0.03. Such information could be highly valuable in inferring the actual operating states of complex photonic systems and providing instruction for achieving the design targets. For example, if the objective is to design a Butterworth filter, we could straightforwardly conclude that both coupling coefficients need to be slightly increased (ideally: $\kappa_r$=0.33 and $\kappa_b$=0.04).

Building upon the broadband reconfigurability of our flat-top filter, we demonstrate ultra-wide measurement bandwidth up to tens of THz by stitching a series of 40-GHz measurement channels, as the schematic diagrams in Figure 3e illustrate. Within each 120-GHz spectral period of the flat-top filter, e.g. between $f_{ci}$ and $f_{c(i+1)}$, we strategically perform four measurements by biasing the filter at relative MZI phases of 0, $\pi/2$, $\pi$, and $3\pi/2$, leading to a frequency offset of 30 GHz between adjacent measurement channels. Considering each measurement could yield a reliable measurement bandwidth of 40 GHz (between $f_c$+10 GHz and $f_c$+50 GHz) with a near-flat spectral response and high SSR, this provides an effective overlapping region of 10 GHz between adjacent channels that could be used to stitch amplitude and phase information of the DUT over a wide frequency range. Figure 3f shows the measured spectral responses of a single microring resonator from 1500 to 1630 nm (16.2 THz), exhibiting a total of 84 resonance notches with an FSR of 219 GHz. The inset shows a zoom-in view between 1626 and 1629 nm with two adjacent resonance notches of the same optical mode. Through a similar analysis as above, we conclude this resonator is at an over-coupling state with a coupling coefficient of 0.11. It should be noted that currently the absolute frequencies of each measurement channels are not accurately referenced since they are only stitched together numerically. Further calibrating the optical carriers in each channel using a phase-stable fiber cavity or interferometer could enable applications that require broadband absolute wavelength accuracy, such as extraction of high-order dispersion information[15].

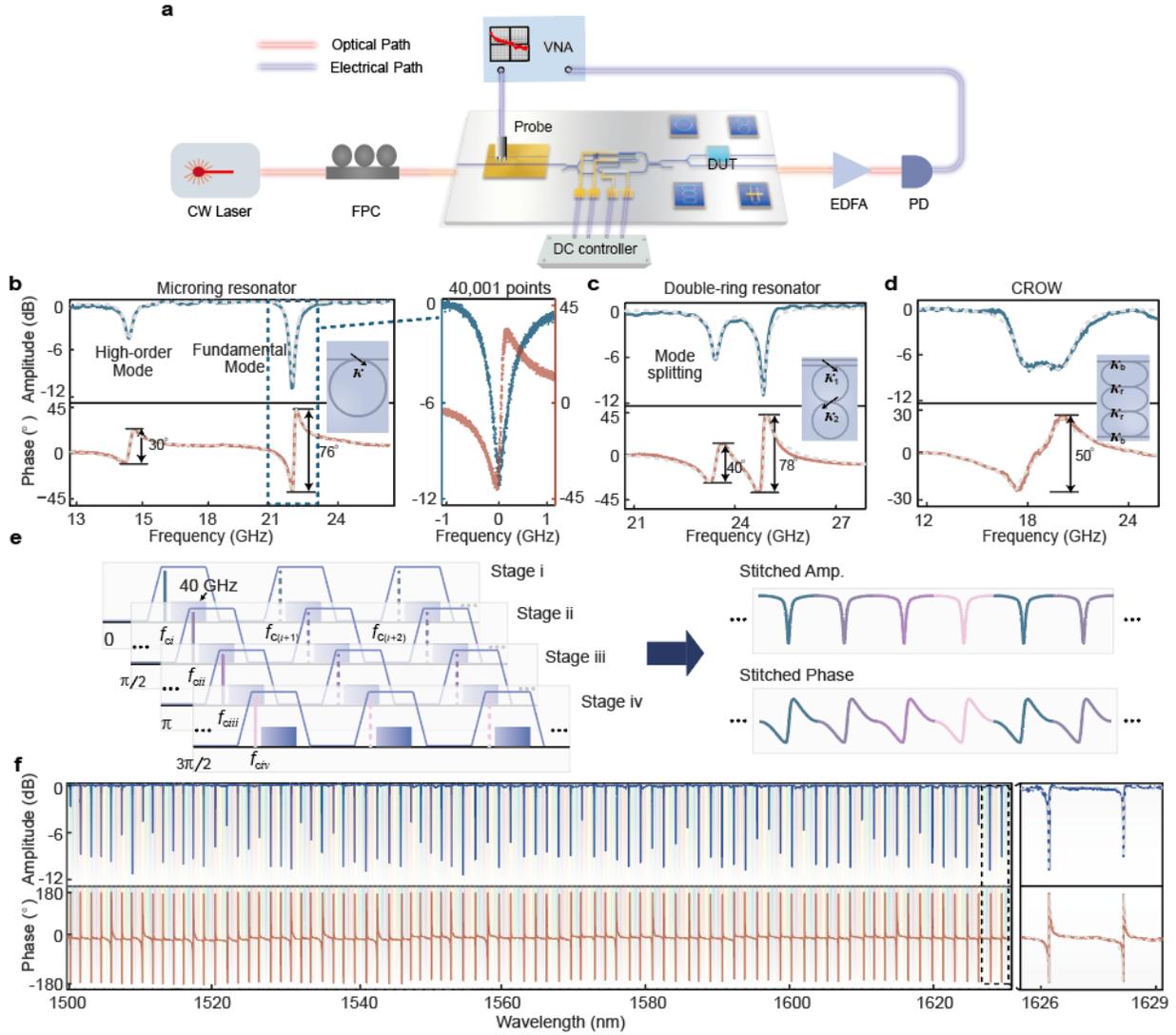

**Fig. 3 In-situ OVA for on-chip passive devices. a** Experimental setup for an in-situ OVA system based on integrated SSB modulation. CW: continuous-wave, FPC: fiber polarization coupler, DUT: devices under test, EDFA: erbium-doped fiber amplifier, PD: photodetector. **b-d** Measured amplitude (blue), phase (red) response, and fitted results (grey dashed) of a single ring resonator (**b**), a double ring system (**c**), and a CROW (**d**). Right panel of (**b**) shows fine measurement results around one resonance dip, with 40001 data points and a resolution of 50 kHz. **e** Working principle of a seamless stitching process for broad measurement bandwidth by tuning the position of the bandpass filter and optical carrier. **f** Ultra-broad-band OVA from 1500 to 1630 nm for a single microring resonator. Inset shows the zoom-in view of the measured responses between 1626 and 1629 nm, indicating an over-coupling state.

4. **In-situ OVA of integrated active devices**

We further demonstrate that our integrated OVA system facilitates the direct characterization of both amplitude and phase evolutions of an actively modulated microring resonator, providing additional probing degrees of freedom for complex physical systems like a synthetic frequency crystal. The concept of frequency synthetic dimension utilizes different optical frequency modes to form a lattice and induce coupling between lattice points through optical nonlinearities [41,42]. Leveraging the large bandwidth, good reconfigurability, and tunable gain/loss of photonic systems, frequency synthetic dimension is well suited to explore complicated systems that are difficult to implement or control in solid states, such as non-Hermitian [43], high-dimensional [44], and topological systems [45]. Here we implement and perform in-situ vector analysis of a synthetic frequency crystal on thin-film LN, which offers the desired low dissipation rate and strong nonlinear interaction via EO effect [44,46,47]. An LN racetrack resonator is employed to generate a set of frequency modes, separated by the FSR of the resonator (24.8 GHz). Efficient modulation, matching the FSR, is achieved through two pairs of electrodes with reversed polarity along the straight sections of the racetrack (Fig. 4a). This creates strong EO coupling between adjacent frequency modes that emulates a tight-binding model [44] (see Methods), where optical photons can hop between different lattice points with a large hopping rate proportional to the microwave driving voltage, which is also known as a resonant EO comb generator [30]. In the steady state of such an actively modulated resonator, input laser signals with different detuning effectively excite different Bloch states whose frequencies match the detuning [Fig. 4b (i) and (ii)]. Traditionally, probing only the optical transmission spectrum reveals the density-of-states (DOS) of the frequency crystal [44]. Here, leveraging the in-situ OVA system, we directly capture both the amplitude and phase response of the Bloch states within the crystal. This direct measurement unveils the collective phase dynamics of the Bloch states [Fig. 4b (iii) and (iv)]. Specifically, for a given laser detuning, a coherent summation [Fig. 4b (iii) to (iv)] of the phase responses of the excited Bloch states [Fig. 4b (i)] is obtained. The reversed phase response between adjacent Bloch states inside the energy band leads to coherent cancellation [Fig. 4b (iii)] on the detected phase dynamics (except for the energy band edges), resulting in a near-flat in-band phase response curve with two sharp phase changes at the top and bottom of the band [Fig. 4b (iv)]. Our experimental results under varying microwave signal powers (0, 15, 18, and 20 dBm) agree well with the theory, as shown in Fig. 4c. By slightly detuning the RF frequency from the resonator FSR (detuning ~ 85 MHz), we further transition the frequency crystal dynamics into a triple resonance state where only three neighboring modes are strongly coupled (Fig. 4d), illustrating the reconfigurability of the amplitude and phase dynamics of the crystal. Our in-situ OVA system represents a stable and compact technology for directly probing active systems with ultrafast modulation, offering extra degrees of freedom for investigating various complex physical systems using frequency synthetic dimensions.

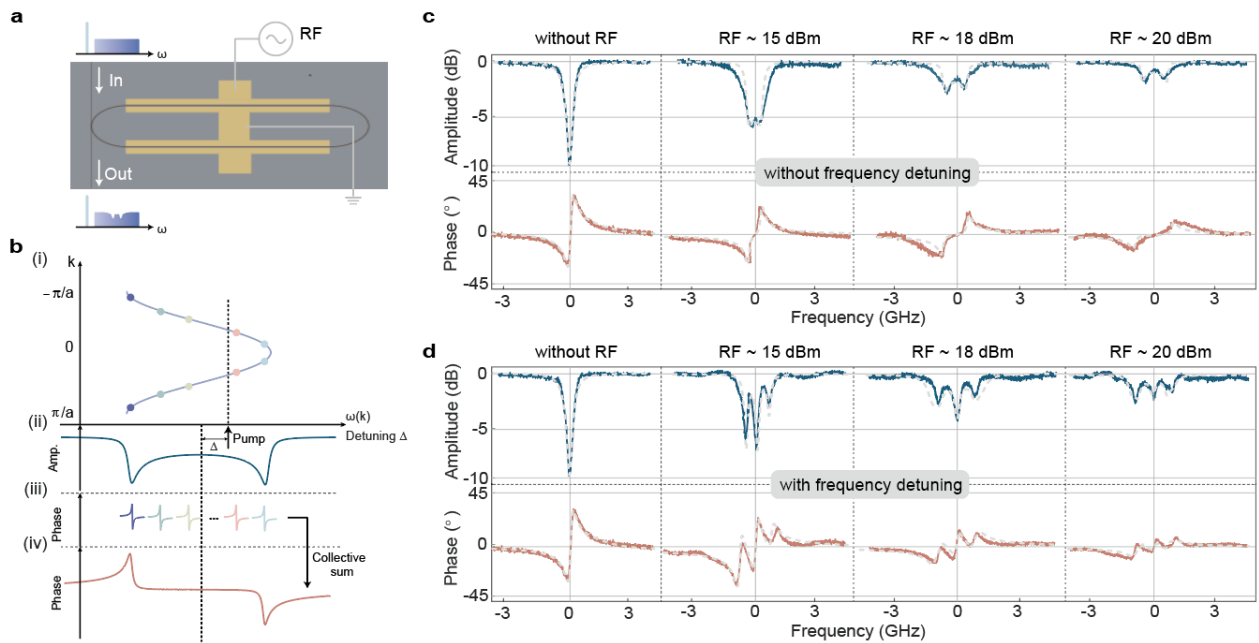

**Fig. 4 In-situ OVA for on-chip active devices. a** Schematic illustration of in-situ probe of an actively modulated microring resonator. Inset shows the microscope image of the LN resonant EO comb generator. **b** (i) Dispersion diagram of the synthetic frequency crystal, where different pump laser detuning excites Bloch modes with different energies and momenta. *a* is the lattice constant. (ii) Density of states of the frequency crystal as a function of laser detuning. (iii-iv) Coherent addition of the phase responses of individual Bloch modes (iii) leads to the measured collective phase dynamics of the Bloch states (iv). **c, d** Measured (solid) and fitted (dashed) amplitude (blue) and phase (red) responses of the synthetic frequency crystal at increasing RF modulation power levels, without (**c**) and with (**d**) frequency detuning. Amp.: amplitude.

## 5. Discussions

In summary, we propose and experimentally implement a miniaturized and flexible OVA system for integrated photonic devices featuring 50 kHz measurement resolution and 16.2 THz measurement bandwidth on the LN platform. The ability to place various DUTs on the same chip as the SSB-based OVA enables in-situ measurements of the amplitude and phase responses, leading to unambiguous and accurate characterization of the intrinsic loss and coupling states of photonic devices. Moreover, we demonstrate the direct probe of full amplitude and phase dynamics of an actively modulated microresonator that emulates the Bloch states in the synthetic frequency dimension. This offers important extra degrees of freedom for understanding and investigating complex physical models like non-Hermitian, high-dimensional, and topological systems.

We envisage that the in-situ OVA could become an important building block in future large-scale PICs that provides accurate and real-time monitoring and feedback control of on-chip elements. This could be achieved by combining the OVA with an electronic field programmable gate arrays (FPGA) chip, which can be trained to process the comprehensive spectral information extracted from the OVA and adjust the operation states of critical PIC components accordingly. Moreover, our demonstrated in-situ OVA system is highly compatible with other high-performance photonic components available on the integrated LN platform, such as frequency comb sources [30,31,48], which could dramatically expedite the processing speed by parallel channelized measurement; and polarization-manipulation components [49,50], which could be adopted to enable the retrieval of polarization dependent loss and polarization group delay. In addition, leveraging the wide transparency window of LN (from visible to mid-infrared) [19], the OVA system could be readily configured to function across a broad range of wavelengths that are of interest to not only optical communications and information technologies, but also biosensing, gas/nanoparticle detection, and atomic physics.

**Methods.**

**Design and fabrication of the devices:**
Devices are fabricated from a commercially available x-cut LNOI wafer (NANOLN), with a 500-nm LN thin film, a 4.7-μm buried $SiO_2$ layer, and a 500-μm silicon substrate. Firstly, $SiO_2$ is deposited on the surface of a 4-inch LNOI wafer as an etching hard mask using plasma-enhanced chemical vapor deposition (PECVD). Various functional devices are patterned on the entire wafer using an ASML UV Stepper lithography system (NFF, HKUST) die-by-die (1.5 cm × 1.5 cm) with a resolution of 500 nm. Next, the exposed resist patterns are transferred first to the $SiO_2$ layer using a standard fluorine-based dry etching process, and then to the LN device layer using an optimized $Ar^+$ based inductively-coupled plasma (ICP) reactive-ion etching process. The LN etch depth is ~ 250 nm, leaving a 250-nm-thick slab. After removal of the residual $SiO_2$ mask and redeposition, an annealing process is carried out. Afterward, a second, third, and fourth lithography and lift-off process is used to fabricate the microwave electrodes, heater, and wires/pads, respectively. Finally, chips are carefully cleaved for end-fire optical coupling with a coupling loss of ~ 5 dB per facet and wire bonded with a printed circuit board (PCB) to realize multi-port control of the TO phase shifters.

The 2×2 MMI is designed through full 3D finite-difference time-domain (FDTD) simulations (Ansys Lumerical) to serve as the resonator coupling region, providing the suitable imparted phase shift. The core MMI region has a size of 31.4 μm × 4.8 μm. The widths of the input waveguides are first tapered from 1.2 μm to 2.15 μm over a taper length of 15 μm before entering the MMI region to minimize scattering loss. The output waveguides are tapered in a similar but reversed manner.

**Principle of RAMZI-based flat-top bandpass filter:**
More details are presented here for the construction of the RAMZI filter. To achieve a "box-like" transfer function instead of the sinusoidal curve of a typical asymmetric MZI structure, the side-coupled ring resonator serves as a periodic phase shifter to adjust the transfer function. The FSR of the microring resonator determines the periodicity of these phase shifts, so the length relationship between the perimeter of the ring resonator and the length imbalance of MZI arms should satisfy $\Delta L_{mzi}=\Delta L_{ring}/2$, where $\Delta L_{mzi}$ is the length difference of the two MZI arms, $\Delta L_{ring}$ is the perimeter of the ring resonator. When the propagation loss is negligible, the optical field of the RAMZI filter can be expressed as:

$$E_{ramzi} = \frac{1}{2}(E_{ring}e^{-i(\phi_1+\phi_{ring})}+1\cdot e^{-i\phi_2}) \tag{1}$$

where $E_{ring}$ is the electric field amplitude of the ring resonator, $\phi_{ring}$ is the phase response of the ring resonator, $\phi_1$ and $\phi_2$ are the phases induced in the two arms of the MZI. The output transfer function of the RAMZI filter can be written as:

$$T_{ramzi} = |E_{ramzi}|^2 = \frac{1}{4}[|E_{ring}|^2+1+2|E_{ring}|\cdot\cos(\phi_{ring}-\phi_{mzi})] \tag{2}$$

where $\phi_{mzi}=\phi_2-\phi_1=\beta\Delta L_{mzi}$, $\beta$ is the propagation constant, and $\phi_{ring}$ can be calculated as:

$$\phi_{ring} = \arctan\frac{(1-\kappa^2)\alpha\sin(\theta)}{(1+\alpha^2)\kappa-\alpha(1+\kappa^2)\cos(\theta)} \tag{3}$$

where $\kappa$ is the optical field coupling coefficient of the ring resonator, $\alpha$ is the amplitude attenuation coefficient of the ring resonator and $\theta$ is the round-trip phase shift. In addition, the RAMZI configuration requires an external phase shift ~ π/2 between two branches of RAMZI to align the resonance notches with the quadrature point of the MZI. Therefore, the phase relation is expressed as $\beta\Delta L_{mzi} = \frac{\beta\Delta L_{mzi}}{2}\pm\frac{\pi}{2}$, and the following relation needs to be satisfied at the off-resonance points [$\theta=(2n+1)\cdot\pi$, n=1,2,3…]:

$$\frac{d\phi_{ring}}{d\theta} = \frac{d\phi_{mzi}}{d\theta} \tag{4}$$

The calculated coupling coefficient of the ring resonator should be $\kappa=\frac{2\sqrt{2}}{3}\approx 0.94$. In this case, the phase response characteristics allow the RAMZI filter to flatten its bandpass region and sharpen the roll-off part.

**Principle of SSB-based OVA system**
More details are presented here for the principle of the SSB-based OVA system. In the analysis of the operation for the SSB-based OVA system, the optical carrier is considered as $e^{j\omega_0 t}$, where $\omega_0$ denotes the carrier frequency. The input RF signal is represented as $\cos\omega_m t$, where $\omega_m$ is the scanning modulation frequency. The input electric field can be written as:

$$E_{in} = e^{jw_0 t}e^{j\beta\cos w_m t} \approx jJ_1(\beta)e^{j(w_0-w_m)t} + J_0(\beta)e^{jw_0 t} + jJ_1(\beta)e^{j(w_0+w_m)t} \tag{5}$$

where $\beta$ is the modulation index, and $J_n(\beta)$ (n = 0, 1) is the Bessel function of the first kind. Considering small-signal modulation, the higher-order sidebands are ignored. After the Fourier transform, the input electric field can be written in the frequency domain as:

$$E_{in}(w) = 2\pi jJ_1(\beta)\delta(w-w_0+w_m) + 2\pi J_0(\beta)\delta(w-w_0) + 2\pi jJ_1(\beta)\delta(w-w_0-w_m) \tag{6}$$

When the signal passes through the flat-top bandpass filter, the optical carrier and one of the sidebands are preserved, resulting in SSB modulation. We assume the first term on the right-hand side of Eq. (6) is suppressed. When the remaining SSB signal is transmitted through the DUT, the output optical signal can be written as:

$$\begin{aligned}E_{out}(w) &= E_{in}(w)H_{sys}(w)H_{DUT}(w) \\ &= 2\pi J_0(\beta)\delta(w-w_0)H_{sys}(w_0)H_{DUT}(w_0) + 2\pi jJ_1(\beta)\delta(w-w_0-w_m)H_{sys}(w_0+w_m)H_{DUT}(w_0+w_m)\end{aligned} \tag{7}$$

where $H_{sys}(\omega)$ and $H_{DUT}(\omega)$ are the transfer functions of the reference arm (used as OVA calibration background) and the DUT, respectively. After sending the signal to an AC-coupled PD, ignoring the DC and high-frequency terms, the output PD current could then be written as:

$$i(w_m) \propto 4\pi^2 jRJ_0(\beta)J_1(\beta)H_{sys}^*(w_0)H_{DUT}^*(w_0)H_{sys}(w_0+w_m)H_{DUT}(w_0+w_m) \tag{8}$$

where $R$ is the responsivity of the PD. The calibration process is performed by measuring the reference arm using the OVA system (without DUT). In this case, it is equivalent to $H_{DUT}(\omega) = 1$, and the output current of the PD could then be written as:

$$i_{sys}(w_m) = 4\pi^2 jRJ_0(\beta)J_1(\beta)H_{sys}^*(w_0)H_{sys}(w_0+w_m) \tag{9}$$

Based on Eq. (8) and (9), the spectral transfer function of the DUT can be obtained by:

$$H(w_m) = H_{DUT}(w_0+w_m) = \frac{i(w_m)}{i_{sys}(w_m)H_{DUT}^*(w_0)} \tag{10}$$

In this equation, $H_{DUT}^*(\omega_0)$ is the spectral response of the DUT at the optical carrier frequency, which can be considered as a complex constant.

**Methodologies of the OVA experiments:**

In our OVA experiment, the optical carrier from the tunable laser (Santec TSL-510) is sent to the LN chip using a lensed fiber after a polarization controller to ensure transverse electric (TE) polarization. The high-resolution scanning RF signals (10~50 GHz) are generated from a vector network analyzer (VNA, E5080B, 50 GHz) and loaded into the modulation electrodes via a high-speed probe (GGB industries, 50 GHz). The output optical signals which contain the full information of DUTs, are amplified using an EDFA (Amonics), detected by a high-speed photodetector (Finisar XPDV21X0RA, 50GHz), and analyzed at another port of VNA. For ultra-wide measurement experiments, two DC sources are utilized to shift the bandpass positions of filters to stitch different measurement channels.

**Fitting of the measured devices parameters:**

In the fitting process for different on-chip devices, initial values and appropriate bounds are set for the parameters to be fitted, and the trust region algorithm is adopted to determine their optimal solution for the theory models. Specifically, for the single microring resonator, the key parameters such as the coupling coefficient ($\kappa$), the resonator round-trip amplitude attenuation factor ($a$), and the perimeter of ring ($L$) are obtained. For the double ring system, building on the directional coupling theory [51], the coupling coefficients of both rings ($\kappa_1$, $\kappa_2$), resonator round-trip amplitude attenuation factors ($a_1$, $a_2$), and the initial resonance frequency difference ($\Delta f$) between two rings can be obtained. For the CROW system, the time-domain coupling model [52] is adopted to fit the curve with the assumption of the identical coupling coefficient $\kappa_r$ in the two ring-ring couplers, identical coupling coefficient $\kappa_b$ in the two ring-bus couplers, as well as identical loss rates of the three rings. Then the coupling coefficients ($\kappa_r$, $\kappa_b$), the resonator round-trip amplitude attenuation factor ($a$), and the perimeter ($L$) are retrieved.

**Principle and simulation of the frequency crystal model:**

The Hamiltonian of the actively modulated microring resonator can be described as [44,46]:

$$H = \sum_{j=-N}^{N} \omega_j a_j^\dagger a_j + \Omega \cos \omega_m t \, (a_j^\dagger a_{j+1} + h.c.) \quad (11)$$

where $\omega_j$ is the frequency of each frequency mode, $\Omega$ is the coupling rate due to modulation, and $\omega_m$ is the frequency of the RF signal. The total number of frequency modes that are coupled are labelled by number -N to N. Using the Heisenberg-Langevin equation, we obtain a set of equations of motion:

$$\dot{a}_j = \left(-i\omega_j - \frac{\kappa}{2}\right) a_j - i\Omega \cos \omega_m t \, (a_{j+1} + a_{j-1}) - \sqrt{\kappa_e} \alpha_{in} e^{-i\omega_L t} \delta_{j,0} \quad (12)$$

Switching the system into the rotating frame by $a_j \to a_j e^{-i\omega_L t} e^{-i j \omega_m t}$ gives:

$$\dot{a}_j = \left(i\Delta + il\delta - \frac{\kappa}{2}\right) a_j - i\frac{\Omega}{2} \left(a_{j+1} e^{i\phi} + a_{j-1} e^{-i\phi}\right) - \sqrt{\kappa_e} \alpha_{in} \delta_{j,0} \quad (13)$$

We perform a transformation: $c_q = \frac{1}{\sqrt{N_t}} \sum_j a_j e^{-iqj}$ and $a_j = \frac{1}{\sqrt{N_t}} \sum_q c_q e^{iqj}$ with $N_t = 2N + 1$. In the case that $\delta = 0$ (note that, for the case of $\delta \neq 0$, it effectively couples less modes therefore the phenomena can be simulated by reducing the $N$), we have

$$\dot{c}_q = \left(i\Delta - \frac{\kappa}{2}\right) c_q - i\Omega \cos(q + \phi) \, c_q - \sqrt{N_t \kappa_e} \alpha_{in} \quad (14)$$

In the steady state, we can derive the expression of $c_q = \sqrt{\kappa_e N_t} \alpha_{in} \frac{1}{i\Delta - \frac{\kappa}{2} - i\Omega \cos(q+\phi)}$. The final output signal from the device can be written as $a_{out} = \left(\alpha_{in} + \sqrt{\kappa_e} \sum_j a_j e^{-ij\omega_m t}\right) = \left(\alpha_{in} + \sqrt{\frac{\kappa_e}{N_t}} c_{q=\omega t}\right)$. The amplitude and phase response can be extracted as $|a_{out}|^2$ and $\text{Arg}(a_{out})$. The photodetector averages the $a_{out}$ within a certain time, therefore the measured phase response is a collective sum of $a_{out}$ over the excited modes $c_q$.

**Funding.** Research Grants Council, University Grants Committee (CityU 11212721, CityU 11204022, N_CityU113/20, C1002-22Y); Croucher Foundation (9509005).

**Acknowledgments.** We thank Prof. Marko Lončar for valuable support and fruitful discussions. We thank the technical support of Mr. Chun Fai Yeung, Mr. Shun Yee Lao, Mr. C W Lai, and Mr. Li Ho at HKUST, Nanosystem Fabrication Facility (NFF) for the stepper lithography and PECVD process. We thank Dr. Wing-Han Wong and Dr. Keeson Shum at CityU for their help in measurement and device fabrication.

**Conflict of interest.** The authors declare no conflicts of interest.

**Data availability.** Data underlying the results presented in this paper are not publicly available at this time but may be obtained from the authors upon reasonable request.

**References**

1 Yang, Z., Albrow-Owen, T., Cai, W. & Hasan, T. Miniaturization of optical spectrometers. *Science* **371**, eabe0722 (2021).
2 Gifford, D. K., Soller, B. J., Wolfe, M. S. & Froggatt, M. E. Optical vector network analyzer for single-scan measurements of loss, group delay, and polarization mode dispersion. *Appl. Opt.* **44**, 7282-7286 (2005).
3 Niemi, T., Uusimaa, M. & Ludvigsen, H. Limitations of phase-shift method in measuring dense group delay ripple of fiber Bragg gratings. *IEEE Photonics Technology Letters* **13**, 1334-1336 (2001).
4 Voges, E., Ostwald, O., Schiek, B. & Neyer, A. Optical phase and amplitude measurement by single sideband homodyne detection. *IEEE J. Quantum Electron.* **18**, 124-129 (1982).
5 Hernandez, R. n., Loayssa, A. & Benito, D. Optical vector network analysis based on single-sideband modulation. *Optical Engineering* **43**, 2418-2421 (2004).
6 Pan, S. & Xue, M. Ultrahigh-resolution optical vector analysis based on optical single-sideband modulation. *J. Lightwave Technol.* **35**, 836-845 (2016).
7 Tang, Z., Pan, S. & Yao, J. A high resolution optical vector network analyzer based on a wideband and wavelength-tunable optical single-sideband modulator. *Opt. Express* **20**, 6555-6560 (2012).
8 Li, W., Wang, W. T., Wang, L. X. & Zhu, N. H. Optical vector network analyzer based on single-sideband modulation and segmental measurement. *IEEE Photonics Journal* **6**, 1-8 (2014).
9 Xue, M., Pan, S., He, C., Guo, R. & Zhao, Y. Wideband optical vector network analyzer based on optical single-sideband modulation and optical frequency comb. *Opt. Lett.* **38**, 4900-4902 (2013).
10 Qing, T. *et al.* Optical vector analysis based on asymmetrical optical double-sideband modulation using a dual-drive dual-parallel Mach-Zehnder modulator. *Opt. Express* **25**, 4665-4671 (2017).
11 Qing, T., Xue, M., Huang, M. & Pan, S. Measurement of optical magnitude response based on double-sideband modulation. *Opt. Lett.* **39**, 6174-6176 (2014).
12 Qing, T., Li, S., Tang, Z., Gao, B. & Pan, S. Optical vector analysis with attometer resolution, 90-dB dynamic range and THz bandwidth. *Nat. Commun.* **10**, 5135 (2019).
13 Li, L. *et al. International Conference on Applied Electromagnetics and Communications.* 1-4 (2016).
14 Chew, S. X. *et al.* Silicon-on-insulator dual-ring notch filter for optical sideband suppression and spectral characterization. *J. Lightwave Technol.* **34**, 4705-4714 (2016).
15 Luo, Y.-H. *et al.* A vector spectrum analyzer of 55.1 THz spectral bandwidth and 99 kHz frequency resolution. *arXiv preprint arXiv:2304.04295* (2023).
16 Marpaung, D., Yao, J. & Capmany, J. Integrated microwave photonics. *Nat. Photon.* **13**, 80-90 (2019).
17 Capmany, J. & Novak, D. Microwave photonics combines two worlds. *Nat. Photon.* **1**, 319 (2007).
18 Feng, H. *et al.* Integrated lithium niobate microwave photonic processing engine. *Nature.* **627**, 80-87 (2024).
19 Zhu, D. *et al.* Integrated photonics on thin-film lithium niobate. *Advances in Optics and Photonics* **13**, 242-352 (2021).
20 Boes, A. *et al.* Lithium niobate photonics: Unlocking the electromagnetic spectrum. *Science* **379**, eabj4396 (2023).
21 Wang, C. *et al.* Integrated lithium niobate electro-optic modulators operating at CMOS-compatible voltages. *Nature* **562**, 101-104 (2018).
22 He, M. *et al.* High-performance hybrid silicon and lithium niobate Mach–Zehnder modulators for 100 Gbit s$^{-1}$ and beyond. *Nat. Photon.* **13**, 359-364 (2019).
23 Feng, H. *et al.* Ultra-high-linearity integrated lithium niobate electro-optic modulators. *Photonics Res.* **10**, 2366-2373 (2022).
24 Xu, M. *et al.* Dual-polarization thin-film lithium niobate in-phase quadrature modulators for terabit-per-second transmission. *Optica* **9**, 61-62 (2022).
25 Kharel, P., Reimer, C., Luke, K., He, L. & Zhang, M. Breaking voltage–bandwidth limits in integrated lithium niobate modulators using micro-structured electrodes. *Optica* **8**, 357-363 (2021).
26 Zhang, M., Wang, C., Cheng, R., Shams-Ansari, A. & Lončar, M. Monolithic ultra-high-Q lithium niobate microring resonator. *Optica* **4**, 1536-1537 (2017).
27 Ke, W. *et al.* Digitally tunable optical delay line based on thin-film lithium niobate featuring high switching speed and low optical loss. *Photonics Res.* **10**, 2575-2583 (2022).
28 Abdelsalam, K. *et al.* Tunable dual-channel ultra-narrowband Bragg grating filter on thin-film lithium niobate. *Opt. Lett.* **46**, 2730-2733 (2021).
29 Li, X. P., Chen, K. X. & Wang, L. F. Compact and electro-optic tunable interleaver in lithium niobate thin film. *Opt. Lett.* **43**, 3610-3613 (2018).
30 Zhang, M. *et al.* Broadband electro-optic frequency comb generation in a lithium niobate microring resonator. *Nature* **568**, 373-377 (2019).
31 Wang, C. *et al.* Monolithic lithium niobate photonic circuits for Kerr frequency comb generation and modulation. *Nat. Commun.* **10**, 978 (2019).
32 Ye, K. *et al.* Surface acoustic wave stimulated Brillouin scattering in thin-film lithium niobate waveguides. *arXiv preprint arXiv:2311.14697* (2023).

33    Rodrigues, C. C. *et al.* On-Chip Backward Stimulated Brillouin Scattering in Lithium Niobate Waveguides. *arXiv preprint arXiv:2311.18135* (2023).

34    Zhu, S. *et al.* Integrated lithium niobate photonic millimeter-wave radar. *arXiv preprint arXiv:2311.09857* (2023).

35    Rizzo, A., Cheng, Q., Daudlin, S. & Bergman, K. Ultra-broadband interleaver for extreme wavelength scaling in silicon photonic links. *IEEE Photonics Technology Letters* **33**, 55-58 (2020).

36    Guan, H. *et al.* Passive silicon ring-assisted Mach–Zehnder interleavers operating in the broadband spectral range. *Appl. Opt.* **59**, 8349-8354 (2020).

37    Song, J. *et al.* Passive ring-assisted Mach-Zehnder interleaver on silicon-on-insulator. *Opt. Express* **16**, 8359-8365 (2008).

38    Peng, B., Özdemir, Ş. K., Chen, W., Nori, F. & Yang, L. What is and what is not electromagnetically induced transparency in whispering-gallery microcavities. *Nat. Commun.* **5**, 5082 (2014).

39    Wang, C., Sweeney, W. R., Stone, A. D. & Yang, L. Coherent perfect absorption at an exceptional point. *Science* **373**, 1261-1265 (2021).

40    Kumar, R. R., Wu, X. & Tsang, H. K. Compact high-extinction tunable CROW filters for integrated quantum photonic circuits. *Opt. Lett.* **45**, 1289-1292 (2020).

41    Yuan, L., Lin, Q., Xiao, M. & Fan, S. Synthetic dimension in photonics. *Optica* **5**, 1396-1405 (2018).

42    Dutt, A. *et al.* Creating boundaries along a synthetic frequency dimension. *Nat. Commun.* **13**, 3377 (2022).

43    Wang, K., Dutt, A., Wojcik, C. C. & Fan, S. Topological complex-energy braiding of non-Hermitian bands. *Nature* **598**, 59-64 (2021).

44    Hu, Y., Reimer, C., Shams-Ansari, A., Zhang, M. & Loncar, M. Realization of high-dimensional frequency crystals in electro-optic microcombs. *Optica* **7**, 1189-1194 (2020).

45    Wang, K. *et al.* Generating arbitrary topological windings of a non-Hermitian band. *Science* **371**, 1240-1245 (2021).

46    Hu, Y. *et al.* Mirror-induced reflection in the frequency domain. *Nat. Commun.* **13**, 6293 (2022).

47    Zhang, K. *et al.* Spectral engineering of optical microresonators in anisotropic lithium niobate crystal. *Adv. Mater.*, 2308840 (2024).

48    Zhang, K. *et al.* A power-efficient integrated lithium niobate electro-optic comb generator. *Communications Physics* **6**, 17 (2023).

49    Chen, Z., Yang, J., Wong, W.-H., Pun, E. Y.-B. & Wang, C. Broadband adiabatic polarization rotator-splitter based on a lithium niobate on insulator platform. *Photonics Res.* **9**, 2319-2324 (2021).

50    Lin, Z. *et al.* High-performance polarization management devices based on thin-film lithium niobate. *Light: Science & Applications* **11**, 93 (2022).

51    Tomita, M., Totsuka, K., Hanamura, R. & Matsumoto, T. Tunable Fano interference effect in coupled-microsphere resonator-induced transparency. *JOSA B* **26**, 813-818 (2009).

52    Liu, H.-C. & Yariv, A. Synthesis of high-order bandpass filters based on coupled-resonator optical waveguides (CROWs). *Opt. Express* **19**, 17653-17668 (2011).